\documentclass[a4paper,12pt]{article}

\usepackage{ifpdf}

\newif\ifpdf
\ifx\pdfoutput\undefined
  \pdffalse
\else
  \pdfoutput=1
  \pdftrue
\fi

\RequirePackage{xspace} %
\RequirePackage{subfigure} %
\RequirePackage[centertags]{amsmath} %
\RequirePackage{amssymb}
\RequirePackage{wrapfig} %
\RequirePackage{calc} %
\RequirePackage{ifthen}
\RequirePackage{tabularx} %
\RequirePackage{flafter} %
\RequirePackage{fancyhdr} %

\ifpdf
  \RequirePackage[pdftex]{color}%
  \RequirePackage{colortbl}%
  \RequirePackage{array}%
  \RequirePackage[pdftex]{graphicx}

  \RequirePackage[ pdftex, plainpages = false, pdfpagelabels,
                 pdfpagelayout = useoutlines,
                 bookmarks,
                 breaklinks = true,
                 linktocpage,
                 pagebackref,                      
                 colorlinks = true,
                 linkcolor = blue,
                 urlcolor  = blue,
                 citecolor = blue,
                 anchorcolor = blue,
                 hyperindex = true,
                 hyperfigures
                 ]{hyperref}

\else
  \RequirePackage{color}
  \RequirePackage{colortbl}
   \RequirePackage{array}
  \RequirePackage[dvips]{graphicx}
  \RequirePackage{hyperref}
  \usepackage{rotating}
\fi

\usepackage{makeidx} 
\usepackage{setspace} 
\usepackage{rotating} 
\usepackage{ecltree}
\usepackage{epic}
\usepackage{supertabular}  
\usepackage{color}
\usepackage{exscale}
\usepackage{fontenc}
\usepackage{ifthen}
\usepackage{latexsym}
\usepackage{makeidx}
\usepackage{syntonly}
\usepackage{inputenc}
\usepackage{graphicx}
\usepackage{setspace}
\usepackage{caption2}
\usepackage[english]{babel}
\usepackage[square, comma,numbers,sort&compress]{natbib}
\usepackage{hypernat}
\usepackage{boxedminipage}
\usepackage{framed}
\usepackage{longtable}
\usepackage[all]{hypcap}    
\usepackage{algorithm2e}
\usepackage{algorithmic}
\usepackage{lscape}
\usepackage{pdflscape}

\newcommand{\note}[1]{\marginpar[left]{\singlespace \tiny #1}}
\newcommand{\pois}{Poiseuille}
\newcommand{\HB}{Herschel-Bulkley}
\newcommand{\pl}{power-law}
\newcommand{\codi}    {converging-diverging} %
\newcommand{\Hs}      {\hspace{-0.5cm}} %
\newcommand{\Vis}    {\mu}            
\newcommand{\lVis}   {\mu_{o}}        
\newcommand{\sR}     {\dot{\gamma}}         
\newcommand{\sS}     {\tau}           
\newcommand{\hsS}    {\sS_{_{1/2}}}   
\newcommand{\wsS}    {\sS_{w}}        
\newcommand{\ysS}    {\sS_{o}}        
\newcommand{\CIF}     {\centering \includegraphics[width=2.7in]} %
\newcommand{\Vmin}    {\vspace{-0.2cm}} %

\renewcommand{\sectionmark}[1]%
      {\markright{\thesection\ #1}} 

\renewcommand{\note}[1]{}

\doublespace 

\title
{ %
\vspace*{3.0cm} \LARGE{\bf Flow of non-Newtonian Fluids in Converging-Diverging Rigid Tubes} \vspace*{4.0cm} \\
}

\author{Taha Sochi\footnote{University College London, Department of Physics \& Astronomy, Gower Street, London, WC1E 6BT.
Email: t.sochi@ucl.ac.uk.} \vspace*{5.0cm}}

\begin{document}

\maketitle %
\pagenumbering{arabic}

\newpage
\phantomsection \addcontentsline{toc}{section}{Contents} %
\tableofcontents

%

\newpage
\phantomsection \addcontentsline{toc}{section}{Abstract} \noindent
{\noindent \LARGE \bf Abstract} \vspace{0.5cm}\\
\noindent %

A residual-based lubrication method is used in this paper to find the flow rate and pressure field
in converging-diverging rigid tubes for the flow of time-independent category of non-Newtonian
fluids. Five converging-diverging prototype geometries were used in this investigation in
conjunction with two fluid models: Ellis and Herschel-Bulkley. The method was validated by
convergence behavior sensibility tests, convergence to analytical solutions for the straight tubes
as special cases for the converging-diverging tubes, convergence to analytical solutions found
earlier for the flow in converging-diverging tubes of Newtonian fluids as special cases for
non-Newtonian, and convergence to analytical solutions found earlier for the flow of power-law
fluids in converging-diverging tubes. A brief investigation was also conducted on a sample of
diverging-converging geometries. The method can in principle be extended to the flow of
viscoelastic and thixotropic/rheopectic fluid categories. The method can also be extended to
geometries varying in size and shape in the flow direction, other than the perfect
cylindrically-symmetric converging-diverging ones, as long as characteristic flow relations
correlating the flow rate to the pressure drop on the discretized elements of the lubrication
approximation can be found. These relations can be analytical, empirical and even numerical and
hence the method has a wide applicability range.

Keywords: fluid mechanics; non-Newtonian fluids; Ellis; Herschel-Bulkley; yield-stress; power-law;
Bingham; converging-diverging tubes; diverging-converging tubes; non-linear systems.

\pagestyle{headings} %
\addtolength{\headheight}{+1.6pt}
\lhead[{Chapter \thechapter \thepage}]%
      {{\bfseries\rightmark}}
\rhead[{\bfseries\leftmark}]%
     {{\bfseries\thepage}} 
\headsep = 1.0cm               

\clearpage
\section{Introduction}

The flow of fluids in general and non-Newtonian in particular through conduits with varying cross
sectional size and/or shape is commonplace in natural and industrial flow systems such as
filtration and refinement devices in the chemical industries and biological fluid transportation
networks like blood circulation vessels and air respiration pathways \cite{WilliamsJ1980,
HooperDM1982, IshikawaGOY1998, Mandal2005, ValenciaZGB2006, FisherR2009, MisraM2012,
SochiNonNewtBlood2013}. Modeling such flows is also needed for analyzing fluid transportation
through porous media where the irregularly-shaped pores and throats are usually described by
geometrically-simple flow conduits with varying cross sections in the flow direction
\cite{White1968, ThienGB1985, ThienK1987, JamesTKBP1990, SouvaliotisB1992, RuthM1993,
SouvaliotisB1996, GrayM2004, SochiComp2010, SochiFeature2010}. In this respect, pore-scale network
modeling is the main candidate and the most natural approach to accommodate such flow-structure
features \cite{Valvatnethesis2004, Lopezthesis2004, Balhoffthesis2005, SochiThesis2007, XuLJ2011,
ToscoMLS2013}. The effect of varying cross section in the flow direction can be particularly
important for modeling certain flow phenomena such as extensional flows, viscoelastic effects and
yield-stress dynamics which are essential for various scientific and technological purposes like
enhanced oil recovery as well as many other applications \cite{SochiVE2009, FedericoPU2010,
AfsharpoorBBH2012, LiuYW2012, RosalesDPBHe2012, DeanoRPAO2012, WangC2012, CirielloF2012,
ChevalierCCDCe2013, AfsharpoorB2013, BalanThesis2013, FarayolaOA2013}.

Several methods have been used in the past to model and simulate such flows; these include
numerical methods like finite element, finite difference and stochastic algorithms
\cite{ThienGB1985, ThienK1987, BurdetteCAB1989, JamesTKBP1990, SochiConDivElastic2013, FuLS2013a},
as well as analytically-based methods \cite{WilliamsJ1980, HooperDM1982, SochiNewtLub2010,
SochiPower2011, SochiNavier2013}. Most previous attempts employ complex mathematical and
computational techniques, which may be unwanted due to difficulties in implementation and
verification, high computational costs, computational complexities like convergence difficulties,
as well as susceptibility to errors and bugs. Also, many of the previous investigations are
relevant only to special cases of fluid or flow or conduit geometry with a subsequent limited use.
Therefore, developing a simple, general and robust method to investigate such flows is highly
desired.

In this paper we use a residual-based lubrication approximation method to find the flow of
non-Newtonian fluids through rigid tubes with \codi\ and diverging-converging shapes in the axial
direction. The method is based on discretizing the flow conduits in the axial dimension into
ring-like elements on which non-Newtonian characteristic flow relations derived for conduits with a
fixed axial geometry can apply. The formulation in this paper is limited to the time-independent
category of the non-Newtonian fluids; although the method in principle is capable of being extended
to history-dependent fluids, i.e. viscoelastic \cite{SochiVE2009} and thixotropic/rheopectic. Two
widely used non-Newtonian fluid models, Ellis and \HB, are examined in this study as prototypes for
the time-independent models to which this method applies. These two fluids can be used to model
diverse non-Newtonian rheological phenomena such as shear-thinning, shear-thickening, and yield
stress as well as Newtonian flow as a special case. These fluid models are commonly used for
describing the time-independent non-Newtonian rheology especially in the industrial applications
such as polymer manufacturing and processing and oil recovery \cite{BalhoffT2006, BalanBNR2011}.

Although the present paper investigates the flow of non-Newtonian fluids in certain axi-symmetric
geometries with converging and diverging features, the method is more general and can be used with
other geometries whose cross section varies in size and/or shape in the flow direction. The only
condition for the applicability of the method is the availability of analytic, empirical or
numerical \cite{SochiVariational2013} relations that correlate the flow rate in the discretized
elements to the pressure drop across these elements.

\section{Investigated non-Newtonian Fluids}

In this section we present a general background about the bulk rheology and flow relations for the
two investigated non-Newtonian fluids: Ellis and \HB.

\subsection{Ellis Model}

This is a three-parameter model that is widely employed to describe the flow of time-independent
shear-thinning yield-free non-Newtonian fluids. The model is known for being successful in
describing the bulk and {\it in situ} rheologies of a wide range of polymeric solutions. It is
particularly used as a replacement for the \pl\ model due to its superiority in matching
experimental data. The model is characterized by a single low-shear-rate Newtonian plateau with the
absence of a high-shear-rate one and hence it may better match the observations in low and medium
shear-rate regimes than in the high shear-rate regimes.

An attractive feature of the Ellis model is the availability of an analytical expression that links
the flow rate to the pressure drop for the flow in cylindrical constant-radius tubes. This feature
makes rheological modeling with Ellis more accurate and convenient than with some other models,
such as Carreau, which have no such analytical expressions although these models may be as good as
or even superior to Ellis. The advantage of having a closed-form analytical flow characterization
relation is obvious especially for some numerical techniques like pore-scale network modeling.

According to the Ellis model, the fluid viscosity $\Vis$ as a function of the shear-stress is given
by the following expression \cite{SadowskiB1965, Savins1969, BirdbookAH1987, CarreaubookKC1997,
SochiThesis2007, SochiArticle2010}

\begin{equation}
    \Vis = \frac{\lVis}{1+ \left(\frac{\sS}{\hsS} \right)^{\alpha - 1}}
\end{equation}
where $\lVis$ is the zero-shear-rate viscosity, $\sS$ is the shear stress, $\hsS$ is the shear
stress at which $\Vis=\frac{\lVis}{2}$ and $\alpha$ is a flow behavior index in this model.

For Ellis fluids, the volumetric flow rate as a function of the pressure drop in a circular
cylindrical tube with a constant radius over its length assuming a laminar incompressible flow with
no slip at the tube wall \cite{SochiSlip2011} is given by the following equation
\cite{SochiThesis2007, SochiB2008}
\begin{equation}\label{QEllis}
    Q = \frac{\pi R^{4} \Delta p}{8 L \lVis}
    \left[ 1 + \frac{4}{\alpha + 3} \left( \frac{R \Delta p}{2 L \hsS}\right)^{\alpha-1} \right]
\end{equation}
where $R$ is the tube radius, $\Delta p$ is the pressure drop over the tube length and $L$ is the
tube length. The derivation of this expression is given in \cite{SochiThesis2007, SochiB2008}.

\subsection{\HB\ Model}

This is a three-parameter model that can describe Newtonian and large group of time-independent
non-Newtonian fluids. According to the \HB\ model, the shear stress as a function of the shear rate
is given by the following relation \cite{Skellandbook1967, SochiThesis2007, SochiArticle2010}
\begin{equation}
    \sS = \ysS + C \sR^{n}
    \verb|       | (\sS > \ysS)
\end{equation}
where $\sS$ is the shear stress, $\ysS$ is the yield-stress above which the substance starts to
flow, $C$ is the consistency coefficient, $\sR$ is the shear rate and $n$ is the flow behavior
index. The \HB\ model reduces to the \pl, or Ostwald-de Waele model, when the yield-stress is zero,
and to the Bingham plastic model when the flow behavior index is unity. It also emulates the
Newtonian fluids when both the yield-stress is zero and the flow behavior index is unity
\cite{LiuM1998}.

The \HB\ model contains six classes: (a) shear-thinning without yield-stress [$n<1.0, \, \ysS=0$]
which is the \pl\ fluid in shear-thinning mode, (b) shear-thinning with yield-stress [$n<1.0, \,
\ysS>0$], (c) Newtonian [$n=1.0, \, \ysS=0$], (d) Bingham plastic [$n=1.0, \, \ysS>0$], (e)
shear-thickening (dilatant) without yield-stress [$n>1.0, \, \ysS=0$], and (f) shear-thickening
with yield-stress [$n>1.0, \, \ysS>0$].

For \HB\ fluids, the volumetric flow rate as a function of the pressure drop in a circular
cylindrical tube with a fixed radius assuming a laminar incompressible flow with no wall slip
\cite{SochiSlip2011, DamianouGM2013} is given by the following relation \cite{Skellandbook1967,
SochiThesis2007, SochiB2008}:

{\footnotesize
\begin{eqnarray} \label{QHerschel}
Q &=& 0 \hspace{10.3cm} (\wsS \le \ysS)
\nonumber \\
    Q &=& \frac{8\pi}{C^\frac{1}{n}}\left(\frac{L}{\Delta p}\right)^{3}
    \left(\wsS - \ysS\right)^{1+\frac{1}{n}}\left[\frac{\left(\wsS - \ysS\right)^{2}}{3+1/n}+
    \frac{2\ysS \left(\wsS - \ysS\right)}{2+1/n} + \frac{\ysS^{2}}{1+1/n}\right]
    \hspace{0.2cm} (\wsS > \ysS)
\end{eqnarray}}
where $\wsS$ is the shear stress at the tube wall which is given by $\wsS=\frac{\Delta pR}{2L}$,
while the meaning of the other symbols have already been given. The derivation of this expression
can be found in \cite{SochiThesis2007, SochiB2008}.

\section{Method} \label{Method}

According to the residual-based lubrication approach, which is proposed in the present paper to
find the flow rate and pressure field in \codi\ tubes for the flow of time-independent
non-Newtonian fluids, the tube is discretized in the axial dimension into ring-like elements. Each
one of these elements is treated as a constant-radius single tube, where the radius of the element
is taken as the average of its inlet and outlet radii, to which Equations \ref{QEllis} and
\ref{QHerschel} apply. From this discretized form of the flow conduit, a system of non-linear
simultaneous equations, which are based on the mass conservation residual plus boundary conditions,
is formed.

For a discretized tube consisting of ($N-1$) elements, there are $N$ nodes: ($N-2$) internal nodes
and two boundary nodes. Each one of these nodes is characterized by a well-defined axial pressure
according to the one-dimensional flow model. Also for the internal nodes, and because of the
assumption of flow incompressibility, the total sum of the volumetric flow rate, which is signed
($+/-$) according to its direction as being toward or away from the given node, is zero due to the
lack of sources and sinks at the node, and hence ($N-2$) residual functions whose essence is the
vanishing of the net flow at the internal nodes are formed. These residual equations are coupled
with two given boundary conditions at the inlet and outlet nodes to form a system of $N$
simultaneous equations.

A typical method for solving such a system is to use an iterative non-linear solution scheme such
as Newton-Raphson method where an initial guess for the internal nodal pressures is proposed and
used in conjunction with the Jacobian matrix of the residual functions to find the pressure
perturbation vector which is then used to adjust the values of the internal nodal pressure. This
iterative process is repeated until a convergence condition, which is usually based on the size of
the residual norm, is reached.

In more formal terms, the process is based on solving the following matrix equation iteratively

\begin{equation}\label{JDpR1}
\mathbf{J}\Delta\mathbf{p}=-\mathbf{r}
\end{equation}
where $\mathbf{J}$ is the Jacobian matrix, $\mathbf{p}$ is the vector of variables which represent
the pressure values at the boundary and internal nodes, and $\mathbf{r}$ is the vector of residuals
which, for the internal nodes, is based on the continuity of the volumetric flow rate as given by

\begin{equation}
f_{j}=\sum_{i=1}^{m}Q_{i}=0
\end{equation}
where $m$ is the number of discretized elements connected to node $j$ which is two in this case,
and $Q_{i}$ is the signed volumetric flow rate in element $i$ as characterized by Equations
\ref{QEllis} and \ref{QHerschel}. Equation \ref{JDpR1} is then solved in each iteration for
$\Delta\mathbf{p}$ which is then used to update $\mathbf{p}$. The convergence will be announced
when the norm of the residual vector, $\mathbf{r}$, becomes within a predefined tolerated marginal
error. More details about this solution scheme can be found in \cite{SochiTechnical1D2013,
SochiPoreScaleElastic2013, SochiConDivElastic2013}.

\clearpage
\section{Implementation and Results}

\begin{figure}
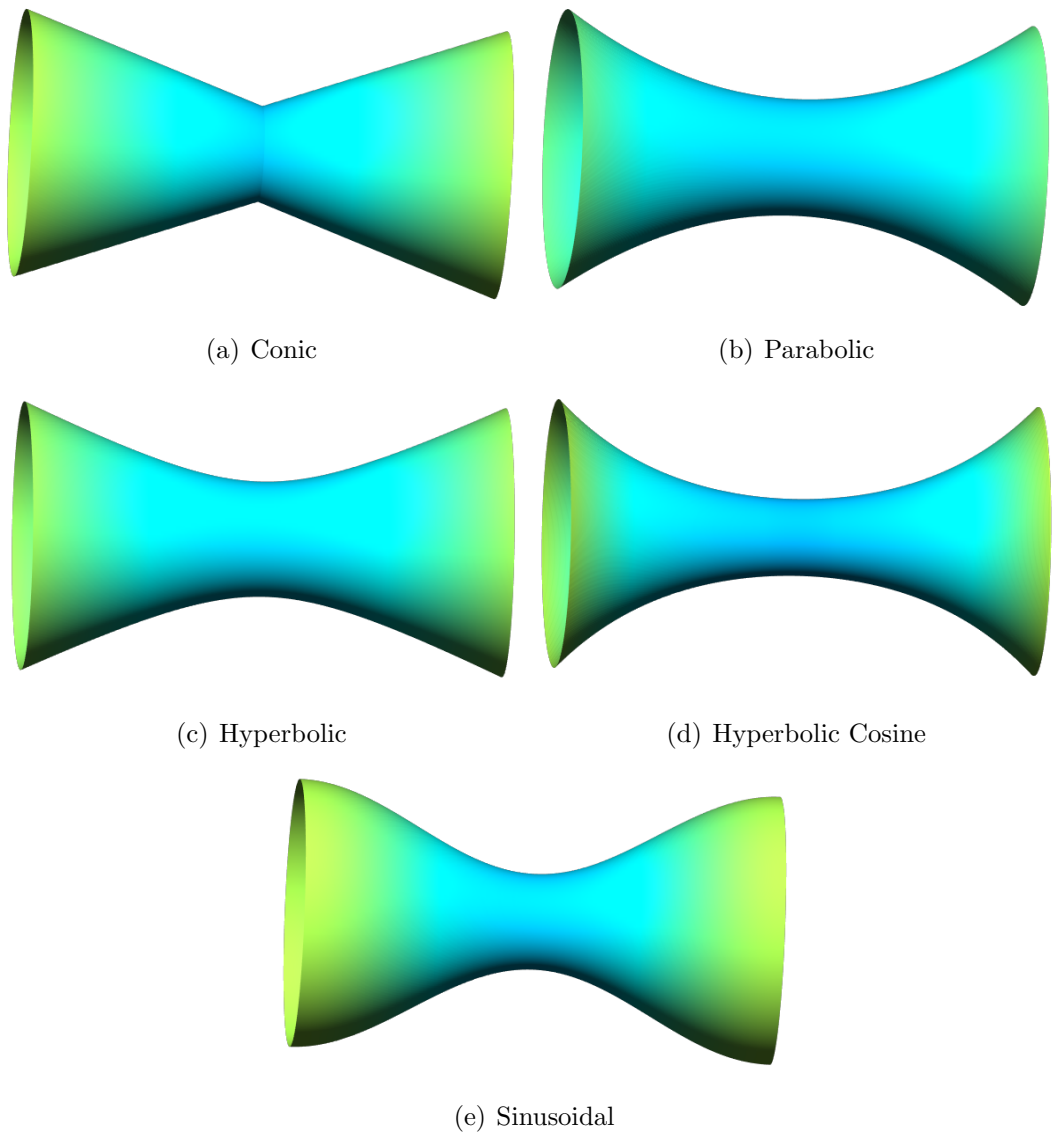

\centering %
\subfigure[Conic]%
{\begin{minipage}[b]{0.5\textwidth} \CIF {g/Conic}
\end{minipage}}
\Hs %
\subfigure[Parabolic]%
{\begin{minipage}[b]{0.5\textwidth} \CIF {g/Parabolic}
\end{minipage}} \Vmin

%
\centering %
\subfigure[Hyperbolic]%
{\begin{minipage}[b]{0.5\textwidth} \CIF {g/Hyperbolic}
\end{minipage}}
\Hs %
\subfigure[Hyperbolic Cosine]%
{\begin{minipage}[b]{0.5\textwidth} \CIF {g/HyperbolicCosine}
\end{minipage}} \Vmin

%
\centering %
\subfigure[Sinusoidal]%
{\begin{minipage}[b]{0.5\textwidth} \CIF {g/Sinusoidal}
\end{minipage}}
\caption{A graphic demonstration of the converging-diverging tube geometries used in this
investigation. \label{Geometries}}
\end{figure}

\begin{figure}[!h]
\centering{}
\includegraphics[scale=0.95]{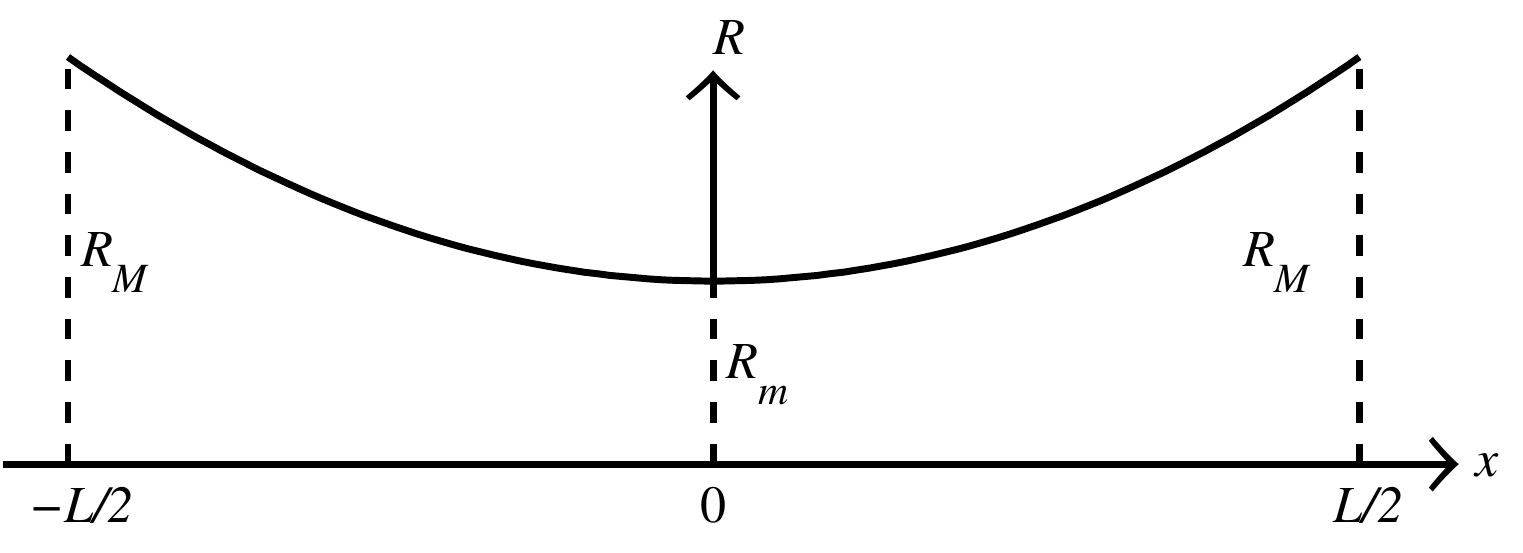}
\caption{A schematic profile of the \codi\ tubes that demonstrates the setting of the coordinate
system used in the equations correlating the tube radius $R$ to its axial coordinate $x$ as used in
Table \ref{GeometryTable}.} \label{GenericProfile}
\end{figure}

The residual-based lubrication method was implemented in a computer code and flow simulation
results were obtained for a wide range of fluid, flow and tube parameters for both Ellis and \HB\
fluids. All the simulations reported in the present paper were carried out using evenly-divided
discretization meshes. In this investigation we used five cylindrically-symmetric \codi\ tube
geometries, a graphic demonstration of which is shown in Figure \ref{Geometries}. The equations
that describe the radius dependency on the tube axial coordinate for these five geometries are
given in Table \ref{GeometryTable}, while a generic \codi\ tube profile, demonstrating the
coordinate system used in the mathematical formulation of this dependency, is shown in Figure
\ref{GenericProfile}.

\begin{figure} [!h]
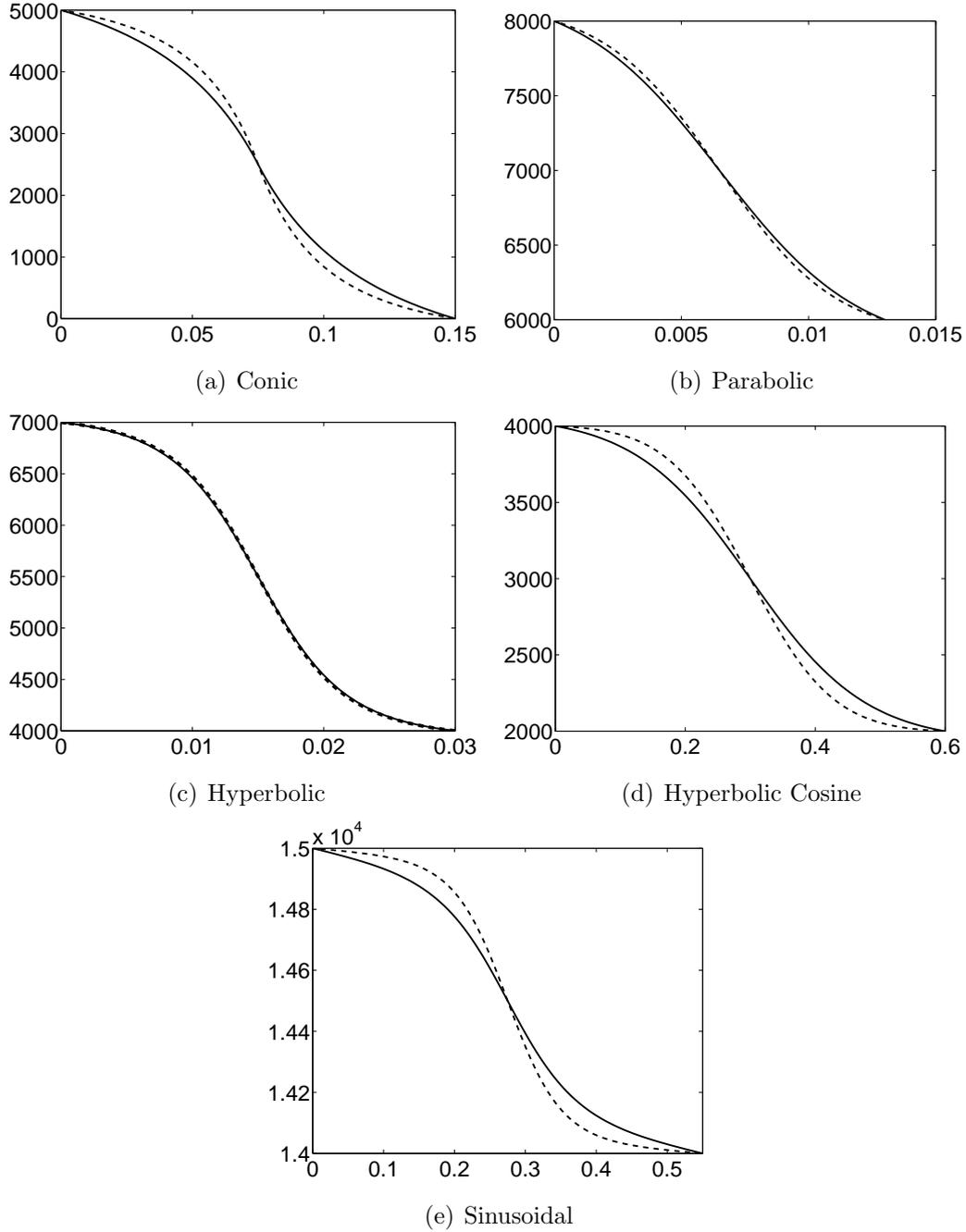

\centering %
\subfigure[Conic]%
{\begin{minipage}[b]{0.5\textwidth} \CIF {g/ConicEllisQ}
\end{minipage}}
\Hs %
\subfigure[Parabolic]%
{\begin{minipage}[b]{0.5\textwidth} \CIF {g/ParabolicEllisQ}
\end{minipage}} \Vmin

%
\centering %
\subfigure[Hyperbolic]%
{\begin{minipage}[b]{0.5\textwidth} \CIF {g/HyperbolicEllisQ}
\end{minipage}}
\Hs %
\subfigure[Hyperbolic Cosine]%
{\begin{minipage}[b]{0.5\textwidth} \CIF {g/CoshEllisQ}
\end{minipage}} \Vmin

%
\centering %
\subfigure[Sinusoidal]%
{\begin{minipage}[b]{0.5\textwidth} \CIF {g/SinusoidalEllisQ}
\end{minipage}}
\caption{Comparing Ellis (solid) to \pois\ (dashed) models for \codi\ rigid tubes of the given
geometry:
(a) conic with $L=0.15$, $R_{m}=0.01$, $R_{M}=0.02$, $p_{i}=5000$, $p_{o}=0$, $Q_E=0.187942$,
$Q_P=0.00452118$
(b) parabolic with $L=0.013$, $R_{m}=0.0017$, $R_{M}=0.0025$, $p_{i}=8000$, $p_{o}=6000$,
$Q_E=3.43037\times 10^{-5}$, $Q_P=1.16241\times 10^{-5}$
(c) hyperbolic with $L=0.03$, $R_{m}=0.002$, $R_{M}=0.004$, $p_{i}=7000$, $p_{o}=4000$,
$Q_E=8.49764\times 10^{-6}$, $Q_P=8.0162\times 10^{-6}$
(d) hyperbolic cosine with $L=0.6$, $R_{m}=0.04$, $R_{M}=0.1$, $p_{i}=4000$, $p_{o}=2000$,
$Q_E=1.8855$, $Q_P=0.00184536$
(e) sinusoidal with $L=0.55$, $R_{m}=0.03$, $R_{M}=0.07$, $p_{i}=15000$, $p_{o}=14000$,
$Q_E=2.14775$, $Q_P=0.00757797$.
All dimensional quantities are in standard SI units. In all five sub-figures, the vertical axis
represents the axial pressure in pascals while the horizontal axis represents the tube axial
coordinate in meters. The rheological properties of the fluids are given in Table
\ref{EllisFluidTable}. \label{EllisPresFig}}
\end{figure}

\begin{figure} [!h]
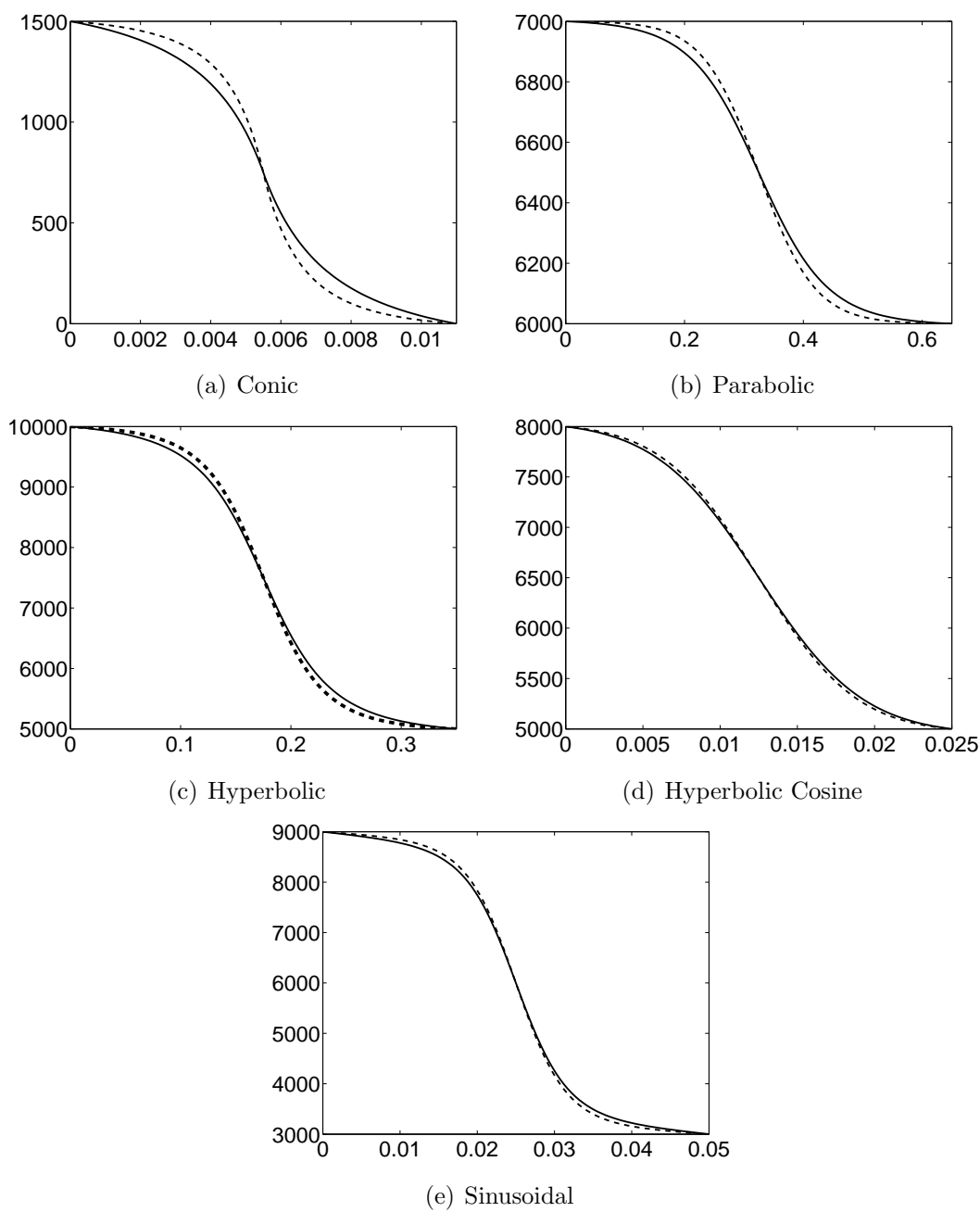

\centering %
\subfigure[Conic]%
{\begin{minipage}[b]{0.5\textwidth} \CIF {g/ConicHBQ}
\end{minipage}}
\Hs %
\subfigure[Parabolic]%
{\begin{minipage}[b]{0.5\textwidth} \CIF {g/ParabolicHBQ}
\end{minipage}} \Vmin

%
\centering %
\subfigure[Hyperbolic]%
{\begin{minipage}[b]{0.5\textwidth} \CIF {g/HyperbolicHBQ}
\end{minipage}}
\Hs %
\subfigure[Hyperbolic Cosine]%
{\begin{minipage}[b]{0.5\textwidth} \CIF {g/CoshHBQ}
\end{minipage}} \Vmin

%
\centering %
\subfigure[Sinusoidal]%
{\begin{minipage}[b]{0.5\textwidth} \CIF {g/SinusoidalHBQ}
\end{minipage}}
\caption{Comparing \HB\ (solid) to \pois\ (dashed) models for \codi\ rigid tubes of the given
geometry:
(a) conic with $L=0.011$, $R_{m}=0.001$, $R_{M}=0.0027$, $p_{i}=1500$, $p_{o}=0$, $Q_H=4.4286\times
10^{-4}$, $Q_P=2.49962\times 10^{-6}$
(b) parabolic with $L=0.65$, $R_{m}=0.04$, $R_{M}=0.15$, $p_{i}=7000$, $p_{o}=6000$, $Q_H=23.9883$,
$Q_P=0.252061$
(c) hyperbolic with $L=0.35$, $R_{m}=0.03$, $R_{M}=0.08$, $p_{i}=10000$, $p_{o}=5000$,
$Q_H=0.0625224$, $Q_P=0.012097$
(d) hyperbolic cosine with $L=0.025$, $R_{m}=0.0025$, $R_{M}=0.005$, $p_{i}=8000$, $p_{o}=5000$,
$Q_H=1.90137\times 10^{-5}$, $Q_P=8.13167\times 10^{-6}$
(e) sinusoidal with $L=0.05$, $R_{m}=0.004$, $R_{M}=0.01$, $p_{i}=9000$, $p_{o}=3000$,
$Q_H=1.84272\times 10^{-4}$, $Q_P=2.04623\times 10^{-4}$.
All dimensional quantities are in standard SI units. In all five sub-figures, the vertical axis
represents the axial pressure in pascals while the horizontal axis represents the tube axial
coordinate in meters. The rheological properties of the fluids are given in Table
\ref{HBFluidTable}. \label{HBPresFig}}
\end{figure}

A sample of the results of these simulations are shown in Figure \ref{EllisPresFig} for Ellis
fluids and Figure \ref{HBPresFig} for \HB\ fluids using the five geometries of Table
\ref{GeometryTable}. In these figures the tube axial pressure is plotted as a function of the tube
axial coordinate for a number of authentic fluids whose bulk rheologies are given in Table
\ref{EllisFluidTable} for the Ellis fluids and Table \ref{HBFluidTable} for the \HB\ fluids. The
pressure dependency for the \pois\ flow of Newtonian fluids having the same zero-shear-rate
viscosity is also given in these figures for demonstration and comparison. The inclusion of the
\pois\ flow can also serve as a sensibility test; for example in Figure \ref{EllisPresFig} (c) we
see a strong similarity in the pressure field and flow rate between the \pois\ and Ellis flows
since the latter has a strong resemblance to the Newtonian at these flow regimes as can be seen
from its rheological parameters. Also for the flow represented by Figure \ref{HBPresFig} (e) we
observe the convergence of the Bingham flow rate to the Newtonian flow rate at this high-pressure
flow regime which is a sensible trend. Unlike the flow rates of the other figures, the
non-Newtonian flow in Figure \ref{HBPresFig} (e) is lower than the Newtonian flow rate because the
non-Newtonian is a Bingham fluid and not a shear-thinning fluid. The effect of the yield-stress in
this case is to lower the flow rate of Bingham from its Newtonian counterpart, as expected. It
should be remarked that in Figures \ref{EllisPresFig} and \ref{HBPresFig}, $p_{i}$ and $p_{o}$
represent the inlet and outlet pressures, while $Q_E$, $Q_H$ and $Q_P$ are the flow rates for
Ellis, \HB\ and \pois\ respectively. The plots shown in these figures represent the converged
solutions which are obtained with the use of discretization meshes that normally consist of 50--100
elements.

\begin{figure}[!h]
\centering{}
\includegraphics[scale=0.55]{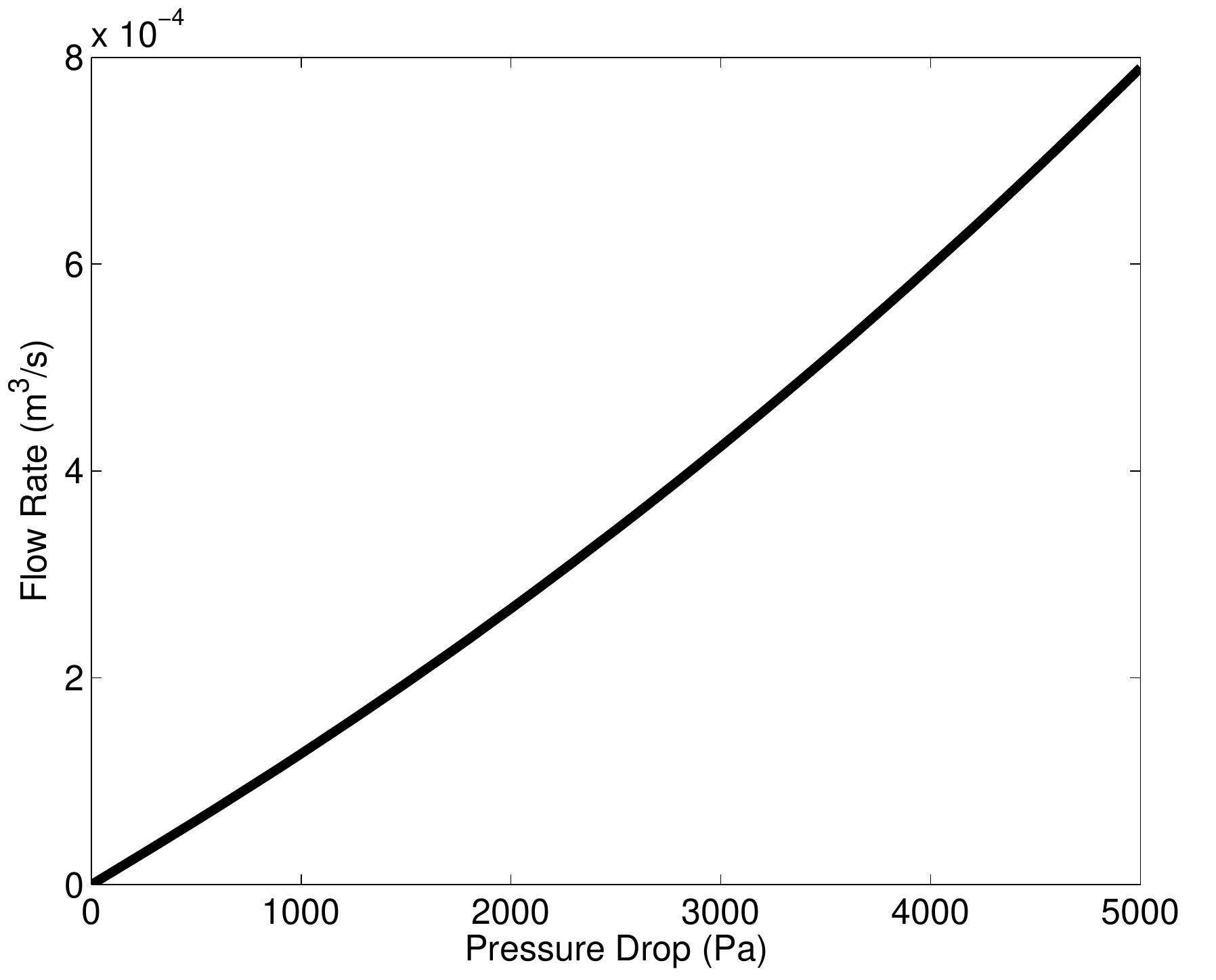}
\caption{Flow rate versus pressure drop for the flow of a 6.0\% Elvanol 72-51 solution modeled by
an Ellis fluid with $\mu_o=0.185$~Pa.s, $\alpha=2.4$ and $\tau_{1/2}=1025$~Pa flowing in a \codi\
conic tube with $L=0.1$~m, $R_{m}=0.005$~m and $R_{M}=0.02$~m. The bulk rheology of the fluid is
taken from Sadowski dissertation \cite{Sadowskithesis1963}.} \label{ConicPQ}
\end{figure}

\begin{figure}[!h]
\centering{}
\includegraphics[scale=0.55]{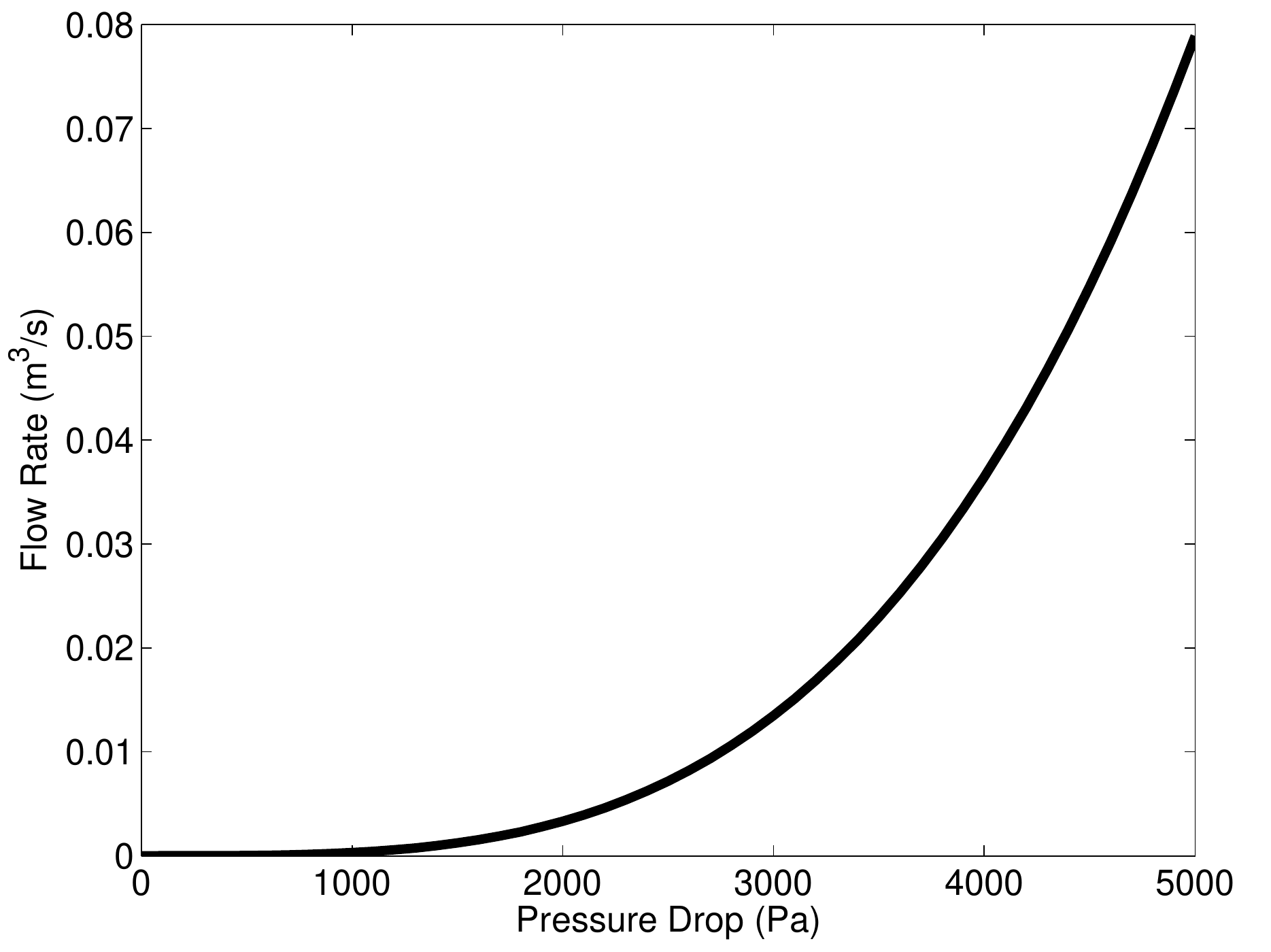}
\caption{Flow rate versus pressure drop for the flow of a guar gum solution of 0.72\% concentration
modeled by an Ellis fluid with $\mu_o=2.672$~Pa.s, $\alpha=3.46$ and $\tau_{1/2}=9.01$~Pa flowing
in a \codi\ parabolic tube with $L=0.06$~m, $R_{m}=0.005$~m and $R_{M}=0.013$~m. The bulk rheology
of the fluid is taken from Balhoff dissertation \cite{Balhoffthesis2005}.} \label{ParabolicPQ}
\end{figure}

\begin{figure}[!h]
\centering{}
\includegraphics[scale=0.55]{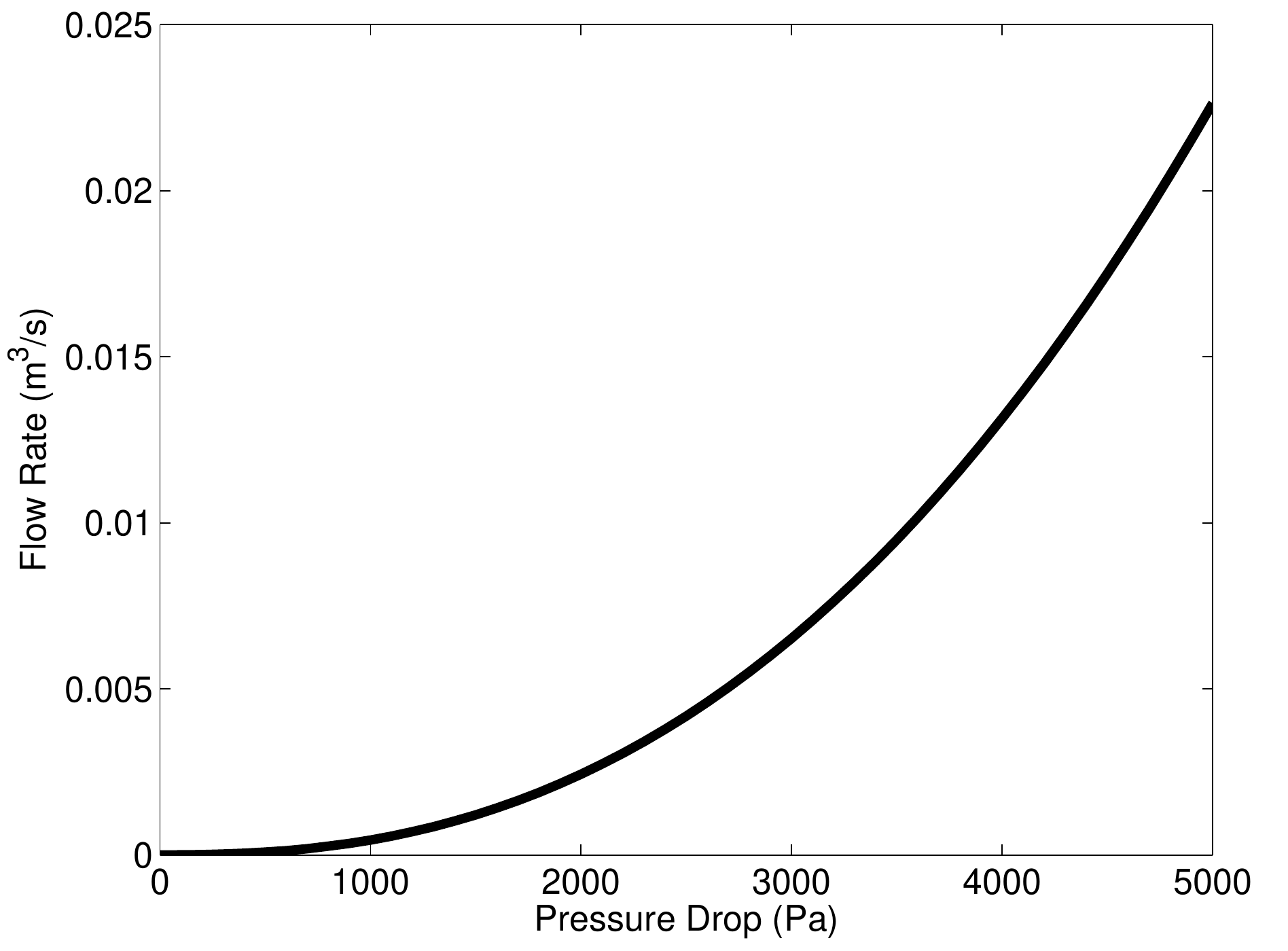}
\caption{Flow rate versus pressure drop for the flow of an aqueous solution of polyacrylamide with
0.25\% concentration modeled by an Ellis fluid with $\mu_o=1.87862$~Pa.s, $\alpha=2.4367$ and
$\tau_{1/2}=0.5310$~Pa flowing in a \codi\ hyperbolic tube with $L=0.025$~m, $R_{m}=0.002$~m and
$R_{M}=0.006$~m. The bulk rheology of the fluid is taken from Park dissertation
\cite{Parkthesis1972}.} \label{HyperbolicPQ}
\end{figure}

\begin{figure}[!h]
\centering{}
\includegraphics[scale=0.55]{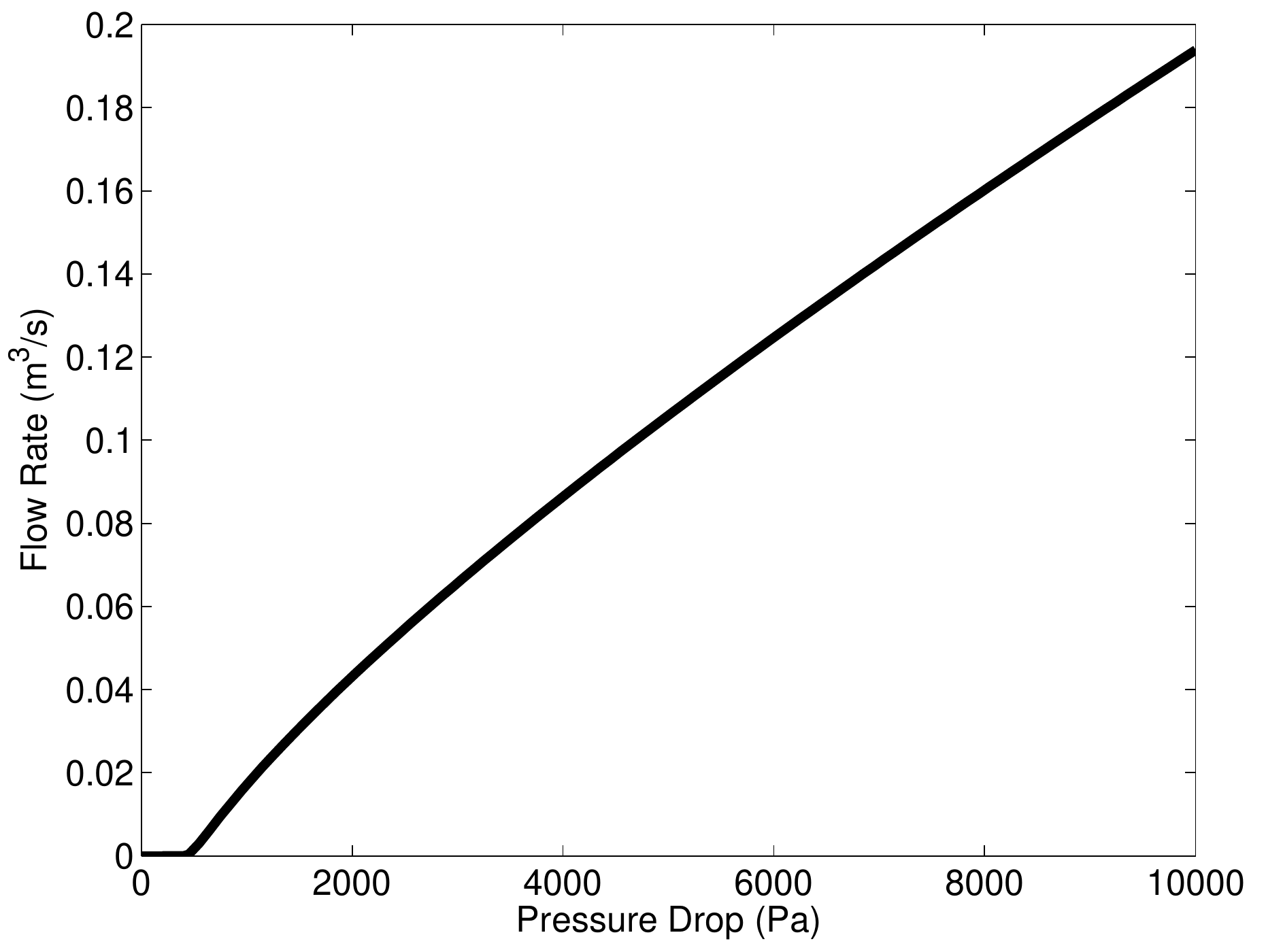}
\caption{Flow rate versus pressure drop for the flow of a hypothetical shear thickening \HB\ fluid
with $C=0.075$~Pa.s$^n$, $n=1.25$ and $\tau_{o}=20.0$~Pa flowing in a \codi\ hyperbolic cosine tube
with $L=0.75$~m, $R_{m}=0.05$~m and $R_{M}=0.15$~m. The threshold yield pressure is about 417~Pa.
Unlike the plots of shear-thinning and Bingham fluids, the curve concave downward due to the
shear-thickening nature.} \label{CoshPQ}
\end{figure}

\begin{figure}[!h]
\centering{}
\includegraphics[scale=0.55]{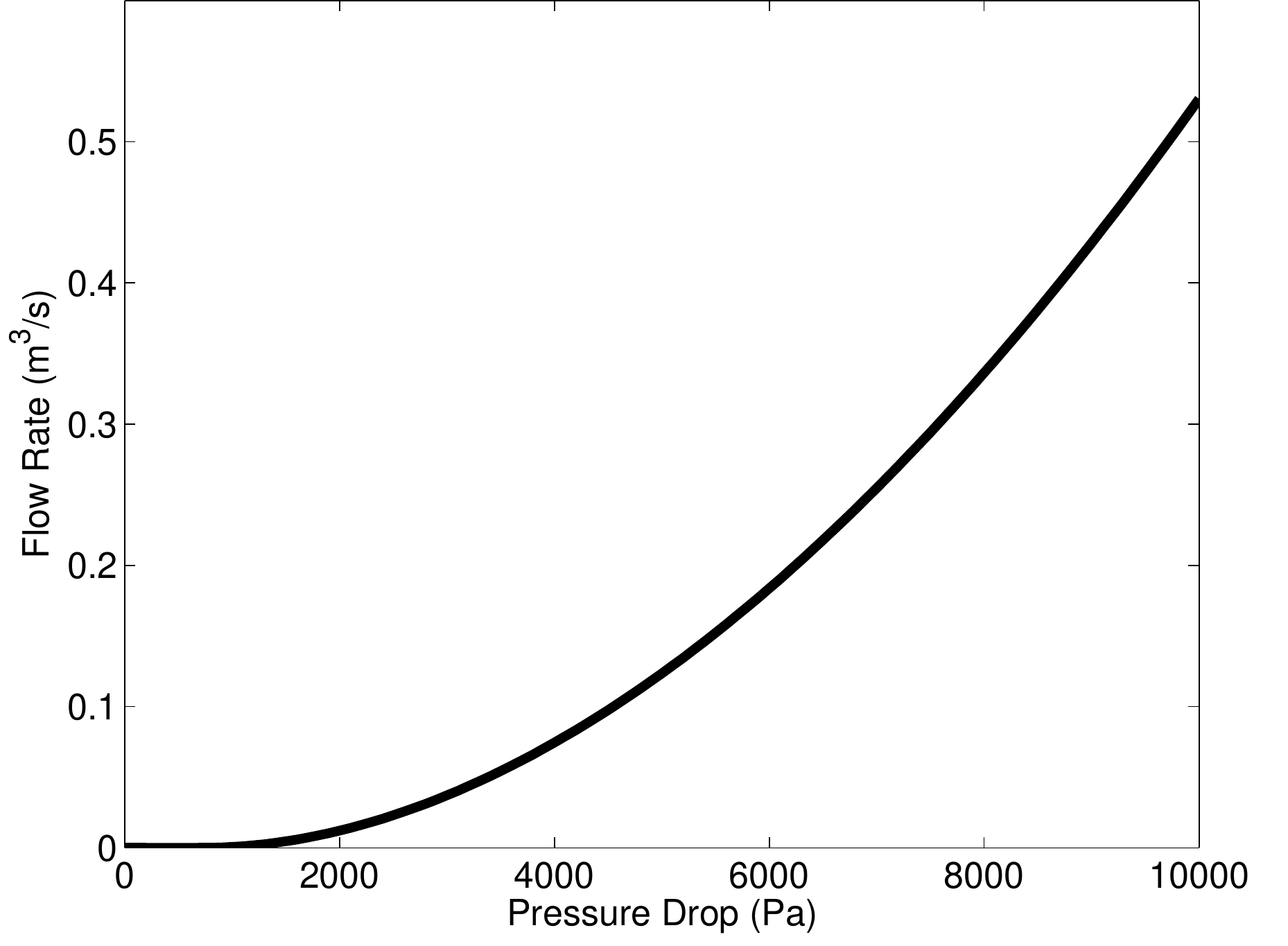}
\caption{Flow rate versus pressure drop for the flow of a hypothetical shear-thinning \HB\ fluid
with $C=0.673$~Pa.s$^n$, $n=0.54$ and $\tau_{o}=20.0$~Pa flowing in a \codi\ sinusoidal tube with
$L=0.5$~m, $R_{m}=0.015$~m and $R_{M}=0.06$~m. The threshold yield pressure is about 664~Pa.}
\label{SinusoidalPQ}
\end{figure}

\begin{figure}
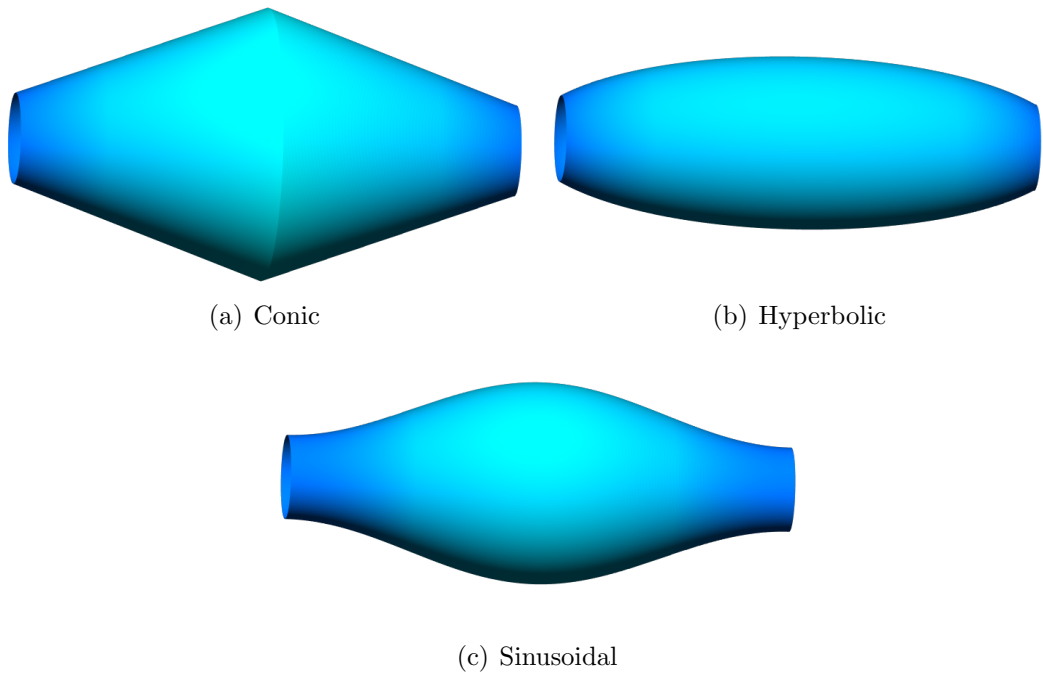

\centering %
\subfigure[Conic]%
{\begin{minipage}[b]{0.5\textwidth} \CIF {g/ConicDC}
\end{minipage}}
\Hs %
\subfigure[Hyperbolic]%
{\begin{minipage}[b]{0.5\textwidth} \CIF {g/HyperbolicDC}
\end{minipage}} \Vmin

%
\centering %
\subfigure[Sinusoidal]%
{\begin{minipage}[b]{0.5\textwidth} \CIF {g/SinusoidalDC}
\end{minipage}}
\caption{A graphic demonstration of a sample of diverging-converging tube geometries.
\label{GeometriesDC}}
\end{figure}

\begin{figure}
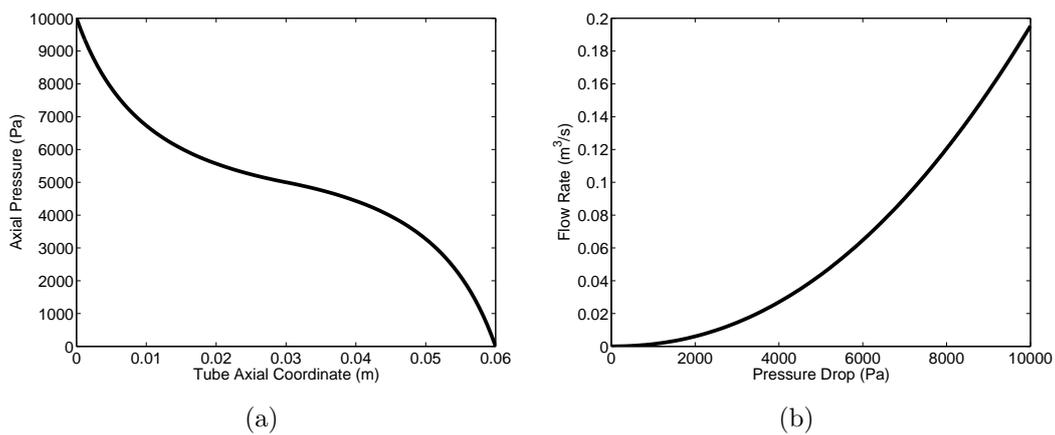

\centering %
\subfigure[]%
{\begin{minipage}[b]{0.5\textwidth} \CIF {g/ConicDCXP}
\end{minipage}}
\Hs %
\subfigure[]%
{\begin{minipage}[b]{0.5\textwidth} \CIF {g/ConicDCPQ}
\end{minipage}}
\caption{Simulated flow of 0.6\% Natrosol - 250H solution modeled by an Ellis fluid with
$\mu_o=0.4$~Pa.s, $\alpha=2.168$ and $\tau_{1/2}=5.2$~Pa flowing in a diverging-converging conic
tube, similar to the one in Figure \ref{GeometriesDC} (a), with $L=0.06$~m, $R_{m}=0.005$~m and
$R_{M}=0.015$~m. The bulk rheology of the fluid is taken from Sadowski dissertation
\cite{Sadowskithesis1963}. \label{ConicFlowDC}}
\end{figure}

\begin{figure}
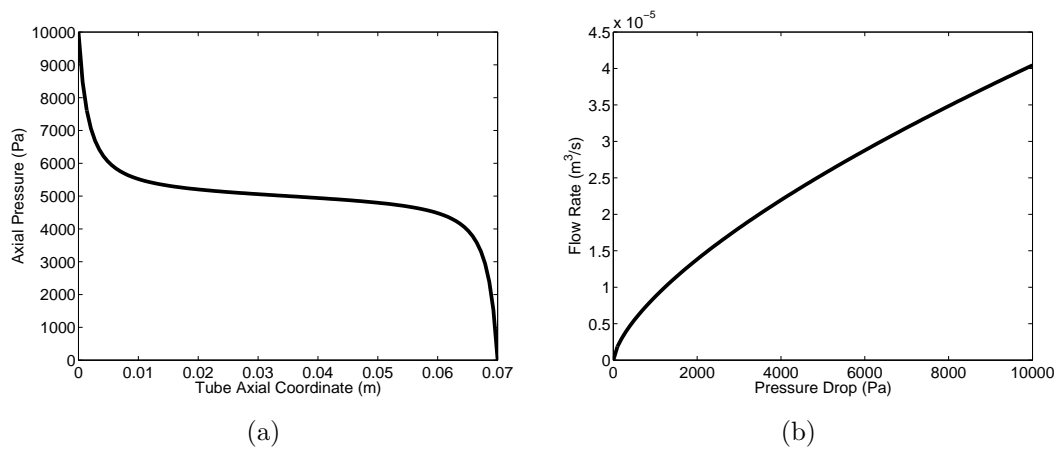

\centering %
\subfigure[]%
{\begin{minipage}[b]{0.5\textwidth} \CIF {g/HyperbolicDCXP}
\end{minipage}}
\Hs %
\subfigure[]%
{\begin{minipage}[b]{0.5\textwidth} \CIF {g/HyperbolicDCPQ}
\end{minipage}}
\caption{Simulated flow of a hypothetical shear-thickening \pl\ fluid with $C=0.75$~Pa.s$^n$ and
$n=1.5$ flowing in a diverging-converging hyperbolic tube, similar to the one in Figure
\ref{GeometriesDC} (b), with $L=0.07$~m, $R_{m}=0.005$~m and $R_{M}=0.013$~m. The flow rate plot
concave downward due to the shear-thickening rheology. \label{HyperbolicFlowDC}}
\end{figure}

\begin{figure}
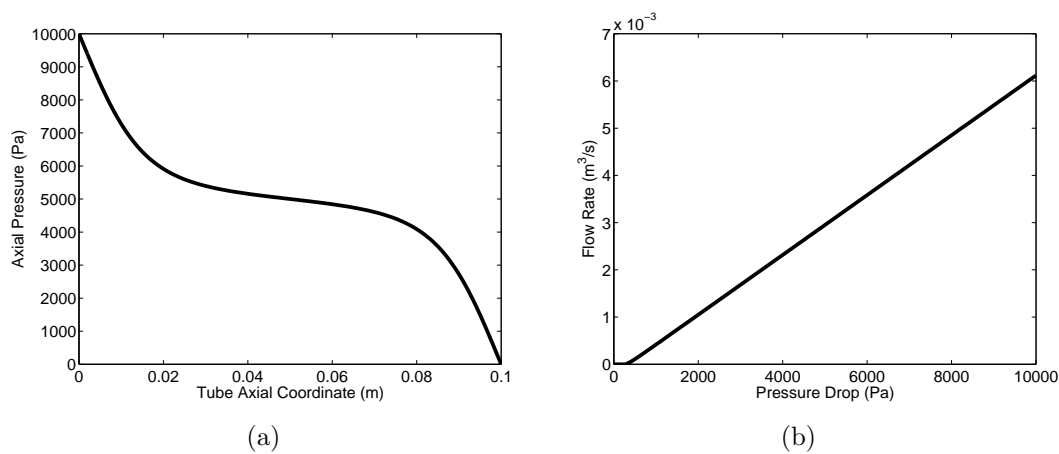

\centering %
\subfigure[]%
{\begin{minipage}[b]{0.5\textwidth} \CIF {g/SinusoidalDCXP}
\end{minipage}}
\Hs %
\subfigure[]%
{\begin{minipage}[b]{0.5\textwidth} \CIF {g/SinusoidalDCPQ}
\end{minipage}}
\caption{Simulated flow of an aqueous solution of Carbopol 941 with 1.0\% concentration modeled by
a Bingham fluid with $C=0.128$~Pa.s and $\tau_{o}=17.33$~Pa flowing in a diverging-converging
sinusoidal tube, similar to the one in Figure \ref{GeometriesDC} (c), with $L=0.1$~m,
$R_{m}=0.009$~m and $R_{M}=0.02$~m. The threshold yield pressure is about 260~Pa. The bulk rheology
of the fluid is taken from Chase and Dachavijit \cite{ChaseD2003}. \label{SinusoidalFlowDC}}
\end{figure}

In Figures \ref{ConicPQ}-\ref{SinusoidalPQ} we present a sample of our results in a different form
where we plot the volumetric flow rate as a function of the pressure drop for a number of \codi\
geometries using a representative sample of Ellis and \HB\ fluids. It should be remarked that the
residual-based lubrication method can also be used with diverging-converging geometries similar to
the ones demonstrated in Figure \ref{GeometriesDC}. A sample of the flow simulation results using
the latter geometries with a representative sample of Ellis and \HB\ fluids is shown in Figures
\ref{ConicFlowDC}-\ref{SinusoidalFlowDC}. In each one of these figures, plots of both axial
pressure versus axial coordinate and flow rate versus pressure drop are given. An interesting
feature is that the curvature of the pressure field plots in these figures is qualitatively
different from that seen in the plots of the \codi\ geometries, which is a sensible tendency as it
reflects the nature of the conduit geometry.


\begin{table} [!h]
\caption{The correlation between the tube radius $R$ and its axial coordinate $x$ for the five
\codi\ geometries used in this paper. In these equations, $-\frac{L}{2}\le x\le\frac{L}{2}$ and
$R_{m}<R_{M}$ where $R_{m}$ is the tube minimum radius at $x=0$ and $R_{M}$ is the tube maximum
radius at $x=\pm \frac{L}{2}$ as demonstrated in Figure \ref{GenericProfile}.}
\label{GeometryTable} \vspace{0.2cm} \centering
\begin{tabular}{ll}
 \hline
Geometry & \hspace{1.5cm}$R(x)$ \\
 \hline
 Conic & $R_{m}+\frac{2(R_{M}-R_{m})}{L}|x|$ \vspace{0.3cm} \\
 Parabolic & $R_{m}+\left(\frac{2}{L}\right)^{2}(R_{M}-R_{m})x^{2}$ \vspace{0.3cm} \\
 Hyperbolic & $\sqrt{R_{m}^{2}+\left(\frac{2}{L}\right)^{2}(R_{M}^{2}-R_{m}^{2})x^{2}}$ \vspace{0.3cm} \\
 Hyperbolic Cosine \hspace{4.5cm} & $R_{m}\cosh\left[\frac{2}{L}\mathrm{arccosh}\left(\frac{R_{M}}{R_{m}}\right)x\right]$ \vspace{0.3cm} \\
 Sinusoidal & $\left(\frac{R_{M}+R_{m}}{2}\right)-\left(\frac{R_{M}-R_{m}}{2}\right)\cos\left(\frac{2\pi x}{L}\right)$ \vspace{0.3cm} \\
\hline
\end{tabular}
\end{table}


\begin{table} [!h]
\caption{Bulk rheology data related to the Ellis fluids used to generate the plots in Figure
\ref{EllisPresFig}.} \label{EllisFluidTable} \vspace{0.2cm} \centering
\begin{tabular}{rccccc}
\hline
           &     Source &      Fluid & $\mu_o$~(Pa.s) &  $\alpha$ & $\tau_{1/2}$~(Pa) \\
\hline
Fig. \ref{EllisPresFig} (a) & Sadowski \cite{Sadowskithesis1963} & 0.4\% Natrosol - 250H &     0.1000 &      1.811 &        2.2 \\

Fig. \ref{EllisPresFig} (b) & Sadowski \cite{Sadowskithesis1963} & 1.4\% Natrosol - 250G &     0.0688 &      1.917 &       59.9 \\

Fig. \ref{EllisPresFig} (c) & Sadowski \cite{Sadowskithesis1963} & 6.0\% Elvanol 72-51 &     0.1850 &      2.400 &     1025.0 \\

Fig. \ref{EllisPresFig} (d) & Park \cite{Parkthesis1972} & polyacrylamide 0.50\% &    4.35213 &     2.4712 &     0.7185 \\

Fig. \ref{EllisPresFig} (e) & Park \cite{Parkthesis1972} & polyacrylamide 0.05\% &    0.26026 &     2.1902 &     0.3390 \\
\hline
\end{tabular}
\end{table}


\begin{table} [!h]
\caption{Bulk rheology data related to the \HB\ fluids used to generate the plots in Figure
\ref{HBPresFig}.} \label{HBFluidTable} \vspace{0.2cm} \centering
\begin{tabular}{rccccc}
\hline
           &     Source &      Fluid & $C$~(Pa.s$^n$) &       $n$ & $\tau_{o}$~(Pa) \\
\hline
Fig. \ref{HBPresFig} (a) & Park \cite{Parkthesis1972} & 0.50\% PMC 400 &      0.116 &       0.57 &      0.535 \\

Fig. \ref{HBPresFig} (b) & Park \cite{Parkthesis1972} & 0.50\% PMC 25 &      0.021 &       0.63 &      0.072 \\

Fig. \ref{HBPresFig} (c) & Al-Fariss \& Pinder \cite{AlfarissP1984} & waxy oil 4\% &      1.222 &       0.77 &      3.362 \\

Fig. \ref{HBPresFig} (d) & Al-Fariss \& Pinder \cite{AlfarissP1984} & waxy oil 5\% &      0.463 &       0.87 &      3.575 \\

Fig. \ref{HBPresFig} (e) & Chase \& Dachavijit \cite{ChaseD2003} & Carbopol 941 1.3\% &      0.215 &       1.00 &      28.46 \\
\hline
\end{tabular}
\end{table}


\clearpage
\section{Tests and Validations}

Several consistency and validation tests have been carried out to verify the residual-based
lubrication method and its code. In the following we outline these tests

\begin{figure}[!h]
\centering{}
\includegraphics[scale=0.55]{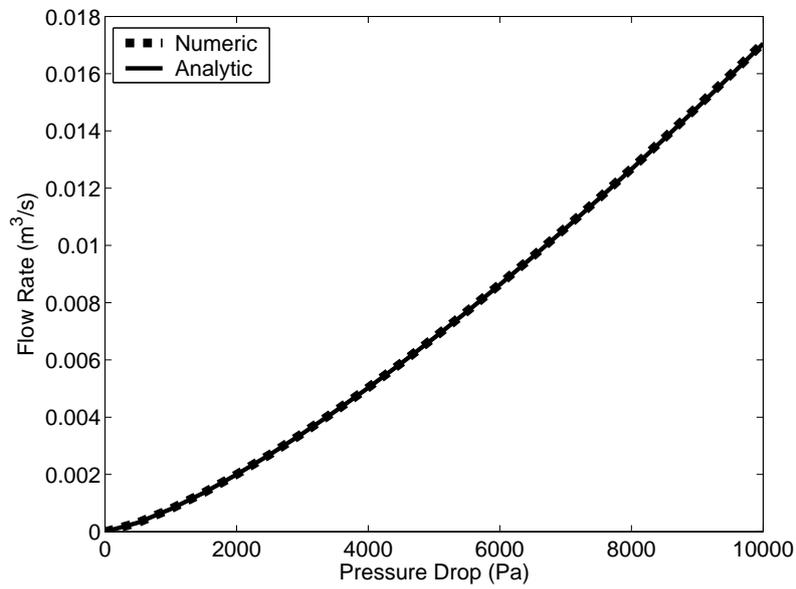}
\caption{Comparing the numerical and analytical solutions for the flow of a typical \pl\ fluid with
$C=0.5$~Pa.s$^n$ and $n=0.75$ through a conically-shaped \codi\ tube with $L=0.15$~m,
$R_{m}=0.01$~m and $R_{M}=0.02$~m. The analytical solution is obtained from the first equation in
Table \ref{QTable}.} \label{ConicPL}
\end{figure}

\begin{itemize}

\item
The sensibility of the method and the integrity of the code have been verified by smooth
convergence to a final solution with mesh refinement by employing finer discretization.

\item
Tests have been carried out on the flow of non-Newtonian fluids through straight cylindrical tubes
with constant radii as a special case for the \codi\ tubes. In all cases the numerical solutions
converged to the correct analytical solutions as given by Equations \ref{QEllis} and
\ref{QHerschel}.

\item
The method and the code have also been verified by the convergence to the correct analytical
solutions for the flow of Newtonian fluids, as a special case for the non-Newtonian fluids, in the
five prototype \codi\ geometries. These geometries have analytical solutions, given in Table
\ref{QTableNewt}, that have been derived and validated previously for the flow of Newtonian fluids
\cite{SochiNewtLub2010, SochiNavier2013, SochiConDivElastic2013}. The Newtonian numerical solutions
have been obtained both as \pois\ flow and as \HB\ flow with $\tau_o=0$ and $n=1$; both of these
numerical solutions were identical to the corresponding analytical solution within acceptable
numerical errors.

\item
The convergence to the correct analytical solution has also been verified for the flow of
non-Newtonian \pl\ fluids through these five \codi\ geometries using analytical expressions that
have been derived and validated previously in \cite{SochiPower2011}; these analytical expressions
are given in Table \ref{QTable}. A sample of these comparisons between the numerical and analytical
solutions is given in Figure \ref{ConicPL} for the flow of a typical \pl\ fluid through a tube with
a conic geometry. As seen, the two solutions are virtually identical within acceptable numerical
errors.

\end{itemize}

\vspace{-0.5cm}


\begin{table} [!h]
\caption{The equations describing the dependency of the flow rate $Q$ on the pressure drop $\Delta
p$ for the flow of Newtonian fluids through rigid tubes with the five \codi\ geometries of Table
\ref{GeometryTable}. These equations were derived and validated previously in
\cite{SochiNavier2013}.}
\label{QTableNewt} \vspace{0.2cm} \centering {
\begin{tabular}{ll}
\hline
Geometry & \hspace{4cm}$Q(\Delta p)$ \vspace{0.2cm} \\
\hline
Conic &  $\frac{3\pi^{2}\Delta p}{\kappa\rho L}\left(\frac{R_{min}^{3}R_{max}^{3}}{R_{min}^{2}+R_{min}R_{max}+R_{max}^{2}}\right)$ \vspace{0.3cm} \\
Parabolic & $\frac{2\pi^{2}\Delta p}{\kappa\rho L}\left(\frac{1}{3R_{min}R_{max}^{3}}+\frac{5}{12R_{min}^{2}R_{max}^{2}}+\frac{5}{8R_{min}^{3}R_{max}}+\frac{5\arctan\left(\sqrt{\frac{R_{max}-R_{min}}{R_{min}}}\right)}{8R_{min}^{7/2}\sqrt{R_{max}-R_{min}}}\right)^{-1}$ \vspace{0.3cm} \\
Hyperbolic & $\frac{2\pi^{2}\Delta p}{\kappa\rho L}\left(\frac{1}{R_{min}^{2}R_{max}^{2}}+\frac{\arctan\left(\sqrt{\frac{R_{max}^{2}-R_{min}^{2}}{R_{min}^{2}}}\right)}{R_{min}^{3}\sqrt{R_{max}^{2}-R_{min}^{2}}}\right)^{-1}$ \vspace{0.3cm} \\
Cosh & $\frac{3\pi^{2}\Delta p}{\kappa\rho L}\left(\frac{\mathrm{arccosh}\left(\frac{R_{max}}{R_{min}}\right)R_{min}^{4}}{\tanh\left(\mathrm{arccosh}\left(\frac{R_{max}}{R_{min}}\right)\right)\left[\mathrm{sech}^{2}\left(\mathrm{arccosh}\left(\frac{R_{max}}{R_{min}}\right)\right)+2\right]}\right)$ \vspace{0.3cm} \\
Sinusoidal & $\frac{16\pi^{2}\Delta p}{\kappa\rho L}\left(\frac{(R_{max}R_{min})^{7/2}}{2(R_{max}+R_{min})^{3}+3(R_{max}+R_{min})(R_{max}-R_{min})^{2}}\right)$ \vspace{0.3cm} \\
\hline
\end{tabular}}
\end{table}

\vspace{-0.7cm}

\begin{table} [!h]
\caption{The equations describing the dependency of the flow rate $Q$ on the pressure drop $\Delta
p$ for the flow of \pl\ fluids in rigid tubes for the five \codi\ geometries of Table
\ref{GeometryTable} where $F_1$ is the Appell hypergeometric function, $_2F_1$ is the
hypergeometric function and $\mathrm{Im}$ is the imaginary part of the given function. These
relations were previously \cite{SochiPower2011} derived and validated.} \label{QTable}
\vspace{0.2cm} \centering {
\begin{tabular}{ll}
 \hline
 Geometry & $ \hspace{3cm} Q(\Delta p)$\tabularnewline
 \hline
 Conic & $\left[\frac{3\pi^{n}n^{n+1}(R_{M}-R_{m})\Delta p}{2LC(3n+1)^{n}}\left(\frac{R_{m}^{3n}R_{M}^{3n}}{R_{M}^{3n}-R_{m}^{3n}}\right)\right]^{1/n}$ \\
 Parabolic & $\left[\frac{n^{n}\pi^{n}R_{m}^{3n+1}\Delta p}{2LC(3n+1)^{n}\,_{2}F_{1}\left(\frac{1}{2},3n+1;\frac{3}{2};1-\frac{R_{M}}{R_{m}}\right)}\right]^{1/n}$ \\
 Hyperbolic & $\left[\frac{\pi^{n}n^{n}R_{m}^{3n+1}\Delta p}{2LC(3n+1)^{n}\,_{2}F_{1}\left(\frac{1}{2},\frac{3n+1}{2};\frac{3}{2};1-\frac{R_{M}^{2}}{R_{m}^{2}}\right)}\right]^{1/n}$ \\
 Hyperbolic Cosine \hspace{3cm} & $\left[\frac{3\pi^{n}n^{n+1}R_{m}R_{M}^{3n}\mathrm{arccosh}\left(\frac{R_{M}}{R_{m}}\right)\Delta p}{2LC(3n+1)^{n}\,\mathrm{Im}\left(_{2}F_{1}\left(\frac{1}{2},-\frac{3n}{2};\frac{2-3n}{2};\frac{R_{M}^{2}}{R_{m}^{2}}\right)\right)}\right]^{1/n}$ \\
 Sinusoidal & $\left[\frac{3\pi^{n+1}n^{n+1}R_{M}^{3n}\sqrt{R_{M}R_{m}}\,\Delta p}{2LC(3n+1)^{n}\mathrm{\, Im}\left(F_{1}\left(-3n;\frac{1}{2},\frac{1}{2};1-3n;1,\frac{R_{M}}{R_{m}}\right)\right)}\right]^{1/n}$ \\
 \hline
\end{tabular}}
\end{table}

\clearpage
\section{Computational Issues and Comparison}

Apart from yield-stress fluids, we did not experience convergence difficulties that normally
associate such non-linear solution schemes like those that we confronted in our previous
investigation to the flow of fluids through distensible structures \cite{SochiPoreScaleElastic2013,
SochiConDivElastic2013}. In almost all the cases investigated in the current study, the convergence
was immediate as it occurred within a few Newton-Raphson iterations even for the high-pressure flow
regime cases with the use of extreme boundary conditions and eccentric parameters for the flow,
fluids and tube geometry.

As for the yield-stress fluids, there were some convergence difficulties and hence to overcome
these difficulties we used a number of numerical tricks. The most effective of these tricks may be
to start with solving the problem assuming zero yield-stress. We then use this yield-free solution
as an initial guess for solving the initial problem with non-zero yield-stress. The final flow
solution in the yield-stress fluid cases is conditioned that if it resulted in $\ysS>\sS_{w}$ on
any element, the flow in the tube is set to zero. It should be remarked that for the flow of
yield-stress fluids through cylindrical tubes with fixed radii, the yield condition where the
pressure drop just overcomes the yield-stress and hence the flow starts is given by the following
relation \cite{SochiThesis2007, SochiB2008, SochiYield2010}

\begin{equation}\label{yieldCondition}
\wsS > \ysS \hspace{1cm} \Longrightarrow \hspace{1cm} \Delta p > {\frac{2 L \ysS}{R}}
\end{equation}

For the high-pressure flow regimes, the convergence is usually more difficult than for the
low-pressure flow regimes, especially for yield-stress fluids. An effective trick in these cases is
to step up on the pressure ladder by starting the process with solving the same problem but with
low pressure boundary conditions where the convergence is easy. This low-pressure solution is then
used as an initial guess for the next step with a higher boundary pressure. On repeating this
process with the use of a suitably divided pressure field, e.g. 100~Pa increase per step,
high-pressure flow regime problems can be solved without convergence difficulties. Although this
stepping up process incurs an extra computational cost, in most cases this extra cost is very low.

Also for the yield-stress fluids, the access of the flow regimes at the border of the threshold
yield pressure, which is required for determining the exact yield point, may not be easy due to the
absence of solutions before yield with possible difficulties in providing a sensible initial guess.
The trick then is to start from a relatively high-pressure flow regime where the convergence to a
solution is easy and where the flow system is expected to have yielded already. The solution at
this regime is then used as an initial guess for stepping down on the pressure ladder, as for
stepping up process which is outlined earlier. By using small pressure steps, e.g. 1~Pa decrease
per step, the flow regimes at the very edge of the threshold yield pressure can be accessed and
hence the exact threshold yield pressure can be determined.

Regarding the CPU time and memory requirements, the residual-based method has a very low
computational cost. This is partly due to the nature of the problem which is related to single
tubes and hence it is normally of limited size. The discretization usually involves a few tens or
at most a few hundreds of elements on the discretization meshes. Hence the computational cost
normally does not exceed a few kilobytes of memory and a few seconds of CPU time on a normal
computer.

The advantages of the residual-based method over the traditional methods is simplicity, generality,
ease of implementation, very good rate and speed of convergence, and very low computational cost.
Moreover, the solutions obtained by the residual-based method match in their accuracy any existing
or anticipated analytical solutions within the given physical and numerical approximations. The
accuracy of the residual-based method and its convergence to the correct analytical solutions are
confirmed in all cases in which analytical solutions are available such as the limiting cases of
straight geometries and Newtonian fluids, as well as the available analytical solutions for the
non-Newtonian \pl\ fluids as given by the equations in Table \ref{QTable}. The main disadvantage of
the residual-based method is its one-dimensional nature and hence it cannot be used for obtaining
the parameters of the flow field in dimensions other than the axial direction. The method has also
other limitations related to the physical assumptions on which the underlying flow model is based.
However, most of these limitations are shared by other comparable methods.

\section{Conclusions}

The investigation in this paper led to the proposal, formulation and validation of a residual-based
lubrication method that can be used to obtain the pressure field and flow rate in rigid tubes with
converging and diverging features, or more generally with a cross section that vary in size and/or
shape in the flow direction, for the flow of non-Newtonian fluids.

The method in its current formulation can be applied to the entire category of time-independent
non-Newtonian fluids, that is generalized Newtonian fluids. Two time-independent non-Newtonian
fluids, namely Ellis and \HB, have been used in this investigation. The method, in our judgement,
has the capability to be extended to the history-dependent non-Newtonian fluids, i.e. viscoelastic
and thixotropic/rheopectic.

Five \codi\ and diverging-converging geometries have been used for test, validation, demonstration
and as prototypes for the investigation of such flows. The method, however, has a wider
applicability range with regard to the tube geometry as it can be applied to all geometries that
vary in size and/or shape in the flow direction as long as a characteristic relation, analytical or
empirical or numerical, that correlates the flow rate to the pressure drop over the
presumably-straight discretized elements can be found.

The residual-based lubrication method was validated by the convergence to the correct analytical
solutions in the special cases of straight constant-radius geometries and Newtonian fluids, as well
as the convergence to the analytical solutions for the flow of \pl\ fluids through the five \codi\
prototype geometries. The method has also been endorsed by a number of qualitatively-sensible
tendencies such as the convergence to a final solution with improved mesh refinement.

The residual-based method has obvious advantages as compared to the more traditional methods. These
advantages include generality, ease of implementation, low computational cost, good rate and speed
of convergence, reliability, and accuracy. The main limitation of the method is its one-dimensional
nature.

\clearpage
\phantomsection \addcontentsline{toc}{section}{Nomenclature} %
{\noindent \LARGE \bf Nomenclature} \vspace{0.5cm}

\begin{supertabular}{ll}
%
$\alpha$             & flow behavior index in Ellis model \\
$\sR$                 & strain rate \\
$\Vis$                & generalized Newtonian viscosity \\
$\lVis$               & zero-shear-rate viscosity \\
$\sS$                 & shear stress \\
$\hsS$                & shear stress when $\Vis = \frac{\lVis}{2}$ in Ellis model \\
$\ysS$                & yield-stress \\
$\wsS$                & shear stress at tube wall \\
\\
$C$                   & consistency coefficient in \HB\ model \\
$f$                     &   flow continuity residual function \\
$F_1$                 & Appell hypergeometric function \\
$_2F_1$               & hypergeometric function \\
$\mathbf{J}$            &   Jacobian matrix \\
$L$                     &   tube length \\
$n$                   & flow behavior index in \HB\ model \\
$N$                     &   number of discretized tube nodes \\
$p$                     &   pressure \\
$\mathbf{p}$            &  pressure vector \\
$p_i$                     &  inlet pressure \\
$p_o$                     &  outlet pressure \\
$\Delta p$              & pressure drop \\
$\Delta\mathbf{p}$      &   pressure perturbation vector \\
$Q$                     &   volumetric flow rate \\
$Q_{E}$                 &   numeric flow rate for Ellis model \\
$Q_{H}$                 &   numeric flow rate for \HB\ model \\
$Q_{P}$                 &   numeric flow rate for \pois\ model \\
$\mathbf{r}$             &  residual vector \\
$R$                     &   tube radius \\
$R_{m}$               & minimum radius of \codi\ tube \\
$R_{M}$               & maximum radius of \codi\ tube \\
$x$                     &   tube axial coordinate \\

\end{supertabular}

\clearpage
\phantomsection \addcontentsline{toc}{section}{References} %
\bibliographystyle{unsrt}

\end{document}